\documentstyle[prb,aps,multicol]{revtex}

\tighten

\begin{document}

\draft


\preprint{\today}

\title{Understanding delocalization in the
Continuous Random Dimer model}

\author{Angel S\'anchez,$^{\dag\ddag}$
Francisco Dom\'{\i}nguez-Adame,$^{\S}$
Gennady Berman,$^{*+}$ and Felix Izrailev$^{\dag\P}$}

\address{$^{\dag}$Theoretical Division and Center for Nonlinear Studies,
Los Alamos National Laboratory, Los Alamos NM 87545\\
$^{\ddag}$Escuela Polit\'ecnica Superior,
Universidad Carlos III de Madrid,
C./ Butarque 15, E-28911 Legan\'es, Madrid, Spain\\
$^{\S}$Departamento de F\'{\i}sica de Materiales,
Facultad de F\'{\i}sicas, Universidad Complutense,
E-28040 Madrid, Spain\\
$^*$Complex Systems Group T-13, Theoretical Division,
Los Alamos National Laboratory, Los Alamos NM 87545\\
$^+$Kirensky Institute of Physics, Krasnoyarsk 660036, Russia\\
$^{\P}$Budker Institute of Nuclear Physics, Novosibirk 630090, Russia}

\maketitle

\begin{abstract}

We propose an explanation of the bands of extended states appearing in
random one dimensional models with correlated disorder, focusing
on the Continuous Random Dimer model [A.\ S\'{a}nchez, E.\ Maci\'a,
and F.\ Dom\'\i nguez-Adame, Phys.\ Rev.\ B {\bf 49}, 147 (1994)].
We show exactly that the transmission coefficient at the resonant
energy is independent of the number of host sites between two consecutive
dimers. This allows us to understand why are there bands of extended states
for every realization of the model as well as the dependence of the bandwidths
on the concentration. We carry out a perturbative calculation that
sheds more light on the above results. In the conclusion we discuss
generalizations of our results to other models and possible applications
which arise from our new insight of this problem.

\end{abstract}

\pacs{PACS numbers: 73.20.Jc, 73.20.Dx, 72.20.$-$i, 85.42.$+$m}

\begin{multicols}{2}
\narrowtext

\section{Introduction}

Since the beginning of this decade, a number of disordered one-dimensional
(1D) models have been proposed \cite{Flores,Dunlap,Wu3} which exhibit
nontrivial extended states. The key ingredient of those models
is {\em correlation}: The defects or impurities are introduced in the host
lattice at random, but always forming pairs, i.e., they never appear isolated.
Further research in this and related models supported this unexpected result,
including other grouping rules aside from pairing.
\cite{Bovier,Wu4,indios,Flores2,JPA,PRBKP,Karmakar}
All those results established on firm grounds the existence of bands of
extended states in this class of models, at least in finite size samples.
However, the {\em reasons} for such
bands to arise remain unclear: To the best of our knowledge, there are only
some perturbative results estimating the number of states\cite{Bovier,Wu4},
and some symmetry conditions for the existence of this kind of resonances.
\cite{Wu4} There is no need to stress the importance of achieving a good
understanding of this delocalization phenomenon, both from theoretical and
applied viewpoints: Such advance will certainly be helpful both to settle
down theoretically its relevance and generality, as well as to design new
devices with specific transmission properties. On the other hand, we concern
ourselves with a model which has been much less studied (indeed, we do not
know of any perturbative calculation or related result regarding it)
and which on the
other hand has specific properties. We address these issues in this
Rapid Communication. To this end, in Sec.\ II we present the specific model
we study, the Continuous Random Dimer model,\cite{JPA,PRBKP} and the
properties we will be dealing with.
In Sec.\ III we show how those features may
be understood in terms of the structure of the transmission coefficients
through lattice segments. Finally, in our conclusions we discuss correlated
disordered models in general in view of our results, as well as possible
applications of this work.

\section{The continuous Random Dimer model}

The Continuous Random Dimer Model (CRDM) was introduced in Refs.\
\onlinecite{JPA} and \onlinecite{PRBKP} and is described by the following
Schr\"odinger equation (we use units such that $\hbar=2m=1$):
\begin{equation}
\label{Schr}
\left[-{d^2\phantom{x}\over dx^2} + \sum_n \lambda_n
\delta(x-n)\right] \psi(x) = E\>\psi(x).
\end{equation}
where $\lambda_n>0$ (the extension of the results to the $\lambda_n<0$ case
is straightforward; besides, the
choice of the sign is irrelevant in most applications, e.g., for superlattices
\cite{Diez2}).  To introduce paired correlated
disorder $\lambda_n$ takes only two values at random,
namely $\lambda$ and $\lambda'$, with the constraint that $\lambda'$
appears only in pairs of neighboring sites (dimer).
This model is related to and inspired by the
(tight-binding) Random Dimer Model (RDM) of Dunlap {\em et al.},
\cite{Dunlap,Wu3} and we believe that our work on the CRDM will provide
also relevant ideas for the RDM. However, there are a number of significant
differences between both models.
First, the CRDM exhibits an infinite number
of resonances and their corresponding
bands of extended states,\cite{JPA,PRBKP}
which makes it interesting from the viewpoint
of applications as there are many options to match the Fermi level.
Second, the fact that the CRDM is continuous and includes
multiple scattering
effects gives it a more realistic character, thus supporting
the possibility of seeing these effects in a variety of
actual physical systems.
Finally, the RDM and the CRDM have different parameters:
Whereas the RDM depends on the
on-site energies $\epsilon_a$ and
$\epsilon_b$ and on the hopping term $V$ (see Ref.\ \onlinecite{Dunlap}),
the CRDM depends on the strengths of the $\delta$ functions and the
intersite distance.

We now summarize the main features of the CRDM.
In Ref.\ \onlinecite{PRBKP} we developed a generalized Poincar\'e map
formalism that allows to map exactly general one-dimensional Schr\"odinger
equations onto discrete equations, for any potential allowed in
quantum mechanics.  In particular, Eq.~(\ref{Schr}) is equivalent to
the discrete map
\begin{equation}
\psi_{n+1}+\psi_{n-1}=2\Omega_n\psi_n,
\label{Poincare}
\end{equation}
where $\psi_n\equiv \psi(n)$ and $\Omega_n\equiv \cos q+(\lambda_n/2q)
\sin q$, with $q=\sqrt{E}$. That formalism allowed us to prove that
there are an infinite number of resonant energies for which the reflection
coefficient of a {\em single} dimer vanishes.\cite{nota}
Resonant energies are given by the conditions $|\Omega|\leq 1$ and
$\Omega'=0$, where $\Omega=\cos q +(\lambda/2q)\sin q$ and $\Omega'$
the same but replacing $\lambda$ by $\lambda'$.
The same result can be generalized in a different manner by using a technique
valid for any equation cast in the form of Eq.\ (\ref{Poincare}) as
explained in Ref.\ \onlinecite{Felix}
We further showed that those resonances survive in the presence of a
{\em finite density} of dimers, i.e., in the CRDM, and moreover, that
they give rise to bands of finite width of truly extended states.
This we established by analyzing several magnitudes, among which we take
here the transmission coefficient as an example of the behavior of the
model. We choose this magnitude because it will subsequently be the
main ingredient for our explanation of delocalization.

An example of the mean behavior of the transmission coefficient around one of
the resonant energies is shown in Fig.\ \ref{trans} for a dimer
concentration $c=0.2$ ($c$ is defined as the ratio between the number of
$\lambda'$ and the total number of $\delta$'s in the lattice);
typical realizations behave in the same way,
the only effect of averaging being to smooth out particular features
of realizations keeping only the main common characteristic, i.e., the
wide transmission peak.  This is the property we want to highlight:
Close to single dimer resonances (in the case of Fig.\ \ref{trans}, the
first one, which occurs at $E_r=3.7626\ldots$ for the chosen parameters
$\lambda=1.0$, $\lambda'=1.5$), there is an interval of energies that
shows also very good transmission properties, similar to those of the
resonant energy.  Most important, such interval has always a finite
width, for all values of dimer concentration, $\lambda$ and $\lambda'$
(provided they satisfy the above conditions), or number of sites in the
lattice.  The peak width depends on the order of the resonance (the
higher the resonance the wider the band of states with transmission
coefficient close to unity) and the concentration of dimers (the larger
the concentration, the narrower the peak, being always of finite width
as already stated). Other magnitudes, such as Landauer resistance or
Lyapunov coefficient behave accordingly.\cite{PRBKP}

After collecting the main {\em facts} about the CRDM and its bands of
extended states, we state what is it that we want to {\em explain:}
First, why are there
intervals of energies for which the transmission is
very close to unity {\em for every realization} of the CRDM? Second,
why does the bandwidth decrease with increasing dimer concentration?
Third, why does the bandwith not vanishes when the dimer concentration
goes to 1? It is clear that if we are able to answer those questions,
we would have understood the physical reasons for the appearance of the
extended bands we are concerned with. This we discuss in detail
in the next section.

\section{Transmission coefficient through lattice segments}

Pursuing answers to the above questions,
we have computed the transmission coefficient of structures
formed by $N$ sites of type $\lambda$ sandwiched either between two
dimers or two single impurities, with all that group embedded in a perfect
infinite chain of $\lambda$ sites. The calculation is once again a transfer
matrix one, which yields an expression that can be evaluated with the help
of a computer. The results, obtained by numerical evaluation of those
exact analytical expressions, are plotted in Fig.\ \ref{f2} for the two dimer
case and in Fig.\ \ref{f3} for the two
single impurities case. Let us begin discussing
the dimer results. It is apparent from Fig.\ \ref{f2} that in all cases
considered,
the transmission coefficient is very close to unity for energies in the
neighborhood of the resonant one. This must be related to the fact that
eigenfunctions corresponding to those energies acquire an
extra phase which will be different from the resonant energy condition
(the change of phase has to be $\pi$), but very close to $\pi$ anyway.
The key point is that for any value of $N$ considered, this interval is
{\em always} located around the resonant energy. This must be compared
to Fig.\ \ref{f3}, where it can be seen that for different $N$ values the
position of the perfectly transmitted energies is also different. Therefore,
we can conclude that the physical reason underlying the existence of
bands of extended states is this overlap of good transmission properties
that happens in the CRDM
{\em forced by the resonant energies} of the dimers.

The above results allows us also to understand why the width of the bands
decreases with increasing concentration but being always finite: Note
for the case with $N=1$ in Fig.\ \ref{f2} that even in this case, the
band shows a non zero extent. It is quite clear that in the high
density limit
most occurrences of the $\lambda$ sites
will be of that type, i.e., one $\lambda$ between two dimers. This is
the case that will then govern the total transmission coefficient of
the chain (obviously, groups of dimers will be perfectly transparent
around the resonant energy as this is placed in the $\lambda'$ band).
We thus see that even in the case when the dimer concentration tends to
unity, the structure of the transmission coefficient for $N=1$ will preserve
the band. Upon decreasing the concentration, those cases will be more and
more rare, and the dominant ones will have larger $N$. Fig.\ \ref{f2}
shows the dramatic increase of the band with increasing $N$, and this
is in perfect agreement with the observations for the dilute chain.
\cite{JPA,PRBKP}

It is also possible to carry out a power expansion in $E-E_r$ of the
transmission coefficient $\tau$ close to the resonant energy, starting
from the above-mentioned transfer-matrix results.  Importantly, the
approach is general for any 1D model, because it can be first cast in
the form of Eq.\ (\ref{Poincare}) (see Ref.\ \onlinecite{PRBKP}) and then
treated within the formalism we describe now. For the sake of brevity,
we skip the general formulae and particularize
Eq.~(\ref{Poincare})
for the system $\lambda_1 = \lambda_2 =
\lambda_{N+3} = \lambda_{N+4} = \lambda'$ and $\lambda_n=\lambda$
otherwise.
We introduce reflection $r$ and transmission $t$
amplitudes through the relationships
\begin{equation}
\label{amps}
\psi_n  = \left\{ \begin{array}{ll} e^{ikn}+
r\>e^{-ikn}, & \mbox{if $n\leq 1$,} \\
t\>e^{ikn}, & \mbox{if $n\geq N+4$,}
\end{array} \right.
\end{equation}
where $\cos k=\Omega$, and we define the promotion matrices
\begin{equation}
P=\left(\begin{array}{cc}2\Omega&-1\\ 1&0\end{array}\right),\ \ \
P'=\left(\begin{array}{cc}2\Omega'&-1\\ 1&0\end{array}\right).
\label{promotion}
\end{equation}
Notice that $P$ and $P'$ are
unimodular. The reflection amplitude can be found
as follows
\begin{equation}
r=e^{ik}\,\frac{T_{11}-T_{22}+T_{12}e^{-ik}-T_{21}e^{ik}}
{T_{21}-T_{12}+T_{22}e^{ik}-T_{11}e^{-ik}},
\label{reflection}
\end{equation}
where $T=(P')^2P^N(P')^2$. Taking into account that $P^N=U_{N-1}(\Omega)
P
-U_{N-2}(\Omega)I_2$, $I_2$ and $U_n$ being the $2\times 2$ unity matrix
and the Chebyshev polynomial of second kind, respectively. Thus, the
matrix elements $T_{ij}$ can be easily written down.
So far, this result is
exact for all energies; since we are interested in those
values of $E$ close to $E_r$, a power expansion leads us after
lengthy but straighforward calculations to the result $\tau=1-|r|^2\sim
1-f(N)(E-E_r)^2$, where $f(N)$ is a known, energy-independent function
expressed in terms of Chebyshev polynomials of second kind; the function
depends only on the number of host $\delta$'s between the two
dimers and is shown in Fig.~\ref{f4}. From this figure we observe that
$f(N)$ oscillates between $\sim 0.0001$ and $\sim 0.01$ but,
what is most important, it has an upper bound. In the most
{\em desfavorable} case, when $f(N)$ reaches a local maximum (e.g.\
$N=13$), $\tau \sim 0.9999$ for $|E-E_r|\sim 0.1$. This means that
there exists an energy range of width $0.2$ for which the transmission
probability is reduced at most $0.01\%$ from unity. In addition, the
above perturbative treatment yields a divergence of the localization
length of the form $\sim (E-E_r)^{-2}$, as found in the RDM,
\cite{Bovier} as well as by different means.\cite{Felix}

\section{Conclusions}

In summary, we have explained the existence of bands of extended states
in the CRDM as arising from the property that transmission
is almost perfect for all kinds of segments in the lattice around the
resonant energy, which is not the case if the impurities are not paired.
It is crucial to notice that this holds for {\em every realization} of
the model.
This explanation also accounts for the dependence of the bandwidth on the
dimer concentration and its finiteness for any such concentration.
We have been able to estimate perturbatively the transmission coefficient
and the divergence of the localization length in that energy interval.
The relevance of this perturbative calculation increases if one realizes
that it is possible to compute the mean transmission coefficient
around the resonance by integrating $f(N)$ with the probability of
finding a segment of $N$ host sites in the infinite lattice.
Furthermore,
we believe that this explanation applies to all models in the same kind
of disordered systems with defect grouping, because the calculations on
those models will be formally very similar, as shown in Ref.\
\onlinecite{Felix}.

We note in closing that
the structure of the transmission
coefficient as depicted in Fig.\ \ref{f2} suggests that it is possible to build
devices with tailored properties by designing an {\em ordered} structure
made up of unit cells formed by
dimers with the appropriate number $N$ of host monomers between
them.
In this context, quantum well superlattices can be a perfect
example of such devices,
as it has been shown\cite{Diez2} that the CRDM can be realized in practice
as a GaAs/AlGaAs system.
On the other hand, there has been recently a significant
increase of interest in
nanotechnological applications of monomolecular assemblies on
solid surfaces.\cite{SAM} Such self-assembled monolayers (SAM) can build up
complicated quasi-1D and 2D structures. One of the important properties of
SAM's is that they can show very different
electron conductivity depending on their
composition and structure.
We hope that the above
resonant mechanism
of appearance of the extended states in correlated disodered models will
be relevant
to understand and design SAM's with the desired conduction properties.

\acknowledgments

A.\ S.\ was partially supported by MEC (Spain)/{}Fulbright,
by DGICyT (Spain) through
project PB92-0248, and by the European Union Human Capital and Mobility
Programme through contract ERBCHRXCT930413.
F.\ D.-A.\ is supported by Universidad Complutense through project
PR161/93-4811.
G.\ P.\ B.\ gratefully acknowledges partial support from Linkage Grant
93-1602 from NATO Special Programme Panel on Nanotechnology.
Work at Los Alamos is
performed under the auspices of the U.S.\ Department of Energy.

\begin{figure}
\caption{Transmission coefficient for the CRDM with a dimer
concentration $c=0.2$.  The $\delta$ function strengths are
$\lambda=1,\>\lambda'=1.5$.  Shown is an average over 100 realizations.
Every realization consists of 15 000 scatterers.  The first allowed band
in the perfect lattice is $[0.921,9.870]$.}
\label{trans}
\end{figure}

\begin{figure}
\caption{Transmission coefficient for
two dimers with $N$ host sites in between,
placed in the middle of an otherwise periodic chain,
for $N=1,2,3,4,5,$ and 6 as indicated in the plot (even $N$,
solid lines, odd $N$, dashed lines).
The $\delta$ function strengths are
$\lambda=1,\>\lambda'=1.5$ as in Fig.\ \protect\ref{trans}.}
\label{f2}
\end{figure}

\begin{figure}
\caption{Transmission coefficient for
two single impurities with $N$ host sites in between,
placed in the middle of an otherwise periodic chain,
for $N=1,2,3,4,5,$ and 6 as indicated in the plot (even $N$,
solid lines, odd $N$, dashed lines).
The $\delta$ function strengths are
$\lambda=1,\>\lambda'=1.5$ as in Fig.\ \protect\ref{trans}.}
\label{f3}
\end{figure}

\begin{figure}
\caption{Plot of the function $f(N)$ is shown, where the
transmission coefficient is $\tau \sim 1- f(N)(E-E_r)^2$ close to
the resonance. See text for details.}
\label{f4}
\end{figure}

\end{multicols}


\begin{references}
\bibitem{nota} In addition, this calculation pinpoints another difference
between RDM and CRDM. In the RDM, the energy is $E=q^2$, $q$ being the
corresponding wavevector, where in the CRDM its r\^ole is played by
$2\cos q$, thus giving rise to more than one band.
\end{references}
\end{document}